\newcommand{\roughly}[1]%
    {{\mathrel{\raise.3ex\hbox{$#1$\kern-.75em\lower1ex\hbox{$\sim$}}}}}

\documentclass[12pt]{article}
\usepackage{paper2e}
\usepackage{epsfig}
\usepackage{latexsym}

\begin{document}
\begin{titlepage}

\preprint{UTAP-580, RESCEU-72/07}

\title{Reheating a multi-throat universe by brane motion}

\author{Shinji Mukohyama}

\address{Department of Physics and 
Research Center for the Early Universe,\\
The University of Tokyo, Tokyo 113-0033, Japan}

\address{Institute for the Physics and Mathematics of the Universe,
 University of Tokyo, Chiba 277-8568, Japan}

\begin{abstract}

 We propose a mechanism of reheating after inflation in multi-throat
 scenarios of warped extra dimensions. Validity of an effective field
 theory on the standard model (SM) brane requires that the position of
 the SM brane during inflation be different from the position after
 inflation. The latter is supposed to be near the tip of the SM
 throat but the former is not. After inflation, when the Hubble
 expansion rate becomes sufficiently low, the SM brane starts moving
 towards the tip and eventually oscillates. The SM fields are excited by
 the brane motion and the universe is reheated. Since interaction
 between the brane position modulus and the SM fields is suppressed only
 by the local string scale, the modulus effectively decays into the SM
 fields. 

\end{abstract}

\end{titlepage}

\section{Introduction}

Our universe appears to have many mass scales, including the Planck
scale, the electroweak scale, the scale of inflation, the scale of
cosmic expansion today, etc. One of the most outstanding problems in
theoretical physics is, hence, to explain how hierarchies among those
scales emerge in a natural way.

The $5$-dimensional braneworld scenario proposed by Randall and Sundrum
generates the large hierarchy between the Planck scale and the
electroweak scale by a warped $5$th
dimension~\cite{Randall:1999ee}. Although their original scenario does 
not recover the ordinary $4$-dimensional Einstein gravity at low energy
on the brane confining the standard model (SM) of particle physics, this
flaw was fixed by inclusion of a moduli stabilization
mechanism~\cite{Goldberger:1999uk}.

Moduli stabilization is important not only for the Randall-Sundrum
scenario but also for any models with extra dimensions. Unless all
moduli are stabilized, the ordinary $4$-dimensional Einstein gravity is
not recovered and models with extra dimensions cannot be compatible
with the universe we observe. (One interesting exception is the
localized gravity~\cite{Randall:1999vf}.)

In type IIB string theory, all complex structure moduli are stabilized
by turning on various fluxes. The geometry of extra dimensions are
inevitably warped by those fluxes and, thus, this setup called warped
flux compactification may be considered as a realization of the
Randall-Sundrum scenario in string theory. A warped region of extra
dimensions is called a warped throat and is supposed to have a regular
tip corresponding to a local minimum of the warp factor. The
$10$-dimensional metric in the warped throat is represented as
%
\begin{equation}
 ds_{10}^2 = h^2ds_4^2+h^{-2}ds_6^2,
  \label{eqn:10d-metric}
\end{equation}
where $ds_4^2=\eta_{\mu\nu}dx^{\mu}dx^{\nu}$ ($\mu,\nu=0,\cdots,3$) is
the $4$-dimensional Minkowski metric, $ds_6^2$ is a Calabi-Yau metric on
a $6$-dimensional compact manifold and the warp factor $h$ depends only
on the internal Calabi-Yau direction. In the construction by Giddings, 
Kachru and Polchinski (GKP)~\cite{Giddings:2001yu}, the warp factor at
the tip of a throat is given by 
%
\begin{equation}
 h_{tip} \simeq \exp\left(-\frac{2\pi K}{3g_sM}\right),
  \label{eqn:warp-factor}
\end{equation}
where $M$ and $K$ are numbers of R-R and NS-NS fluxes, respectively,
$g_s$ is the string coupling, and the warp factor is normalized to order
unity in the unwarped, bulk Calabi-Yau region. As in the Randall-Sundrum
scenario, this exponential warp factor generates a large hierarchy
between the mass scale at the tip and that in the bulk Calabi-Yau
region. For example, if the exponent in the right hand side of 
(\ref{eqn:warp-factor}) is $\simeq -37$ then the local string scale at
the tip is as low as the electroweak scale when the string scale in the
bulk region is Planckian. The Einstein gravity is recovered at low
energy if all other moduli, called K\"ahler moduli, are stabilized by 
non-perturbative effects such as D-instantons and/or gaugino
condensation \'a la Kachru, Kallosh, Linde and
Trivedi~\cite{Kachru:2003aw}.

While those scenarios can address the hierarchy between the Planck scale
and the electroweak scale, in our universe there are more than one
hierarchies. In this respect it is tempting to consider more than one 
throats attached to a bulk Calabi-Yau manifold. Each throat has a
tip corresponding to a local minimum of the warp factor and, thus, a
characteristic mass scale, i.e. the local string scale. For example, a
long throat may accommodate intersecting branes confining the SM fields
and another throat with an intermediate length may accommodate a
brane-antibrane pair to drive inflation~\cite{Kachru:2003sx}. In this
way, the Planck scale in the bulk Calabi-Yau region, the electroweak
scale at the tip of the SM throat and the scale of inflation at the tip
of the inflaton throat can coexist in the geometrical setup. On the
other hand, if both the SM and the inflaton are in a common throat then
one might expect that supersymmetry might help accommodate both
scales. However, this turns out to be rather difficult since the
gravitino tends to be heavier than the Hubble scale during
inflation~\cite{Kallosh:2004yh}, which is set to $H_{inf}\sim
10^{-5}\times\sqrt{\epsilon} M_{Pl}$ by the amplitude of temperature
fluctuations. Therefore, it seems plausible to consider more than one
throats: one throat for the inflaton and another for the SM.

The purpose of this paper is to propose a new mechanism to reheat the SM 
fields after inflation in the multi-throat scenario. The rest of this
paper is organized as follows. In Sec.~\ref{sec:problems} we point out
a couple of problems with multi-throat inflation models. Motivated by
those problems, in Sec.~\ref{sec:scenario} we propose a reheating
scenario. Sec.~\ref{sec:summary} is devoted to a summary of this paper.

\section{Problems with multi-throat inflation model}
\label{sec:problems}

In this section we point out a couple of problems with multi-throat
inflation models.

\subsection{Reheating}
\label{subsec:reheating}

Once we adopt a multi-throat scenario, a natural question arises: ``how 
to reheat the SM fields after inflation?'' One of the reasons why this
is not trivial is that the inflaton and the SM fields are living in
different throats. In this subsection we shall review a known mechanism
in the literature~\cite{Barnaby:2004gg,Kofman:2005yz,Chialva:2005zy} and
point out a problem. (See also \cite{Brodie:2003qv}.)

The end of a brane-antibrane inflation is annihilation of the
brane-antibrane pair. When the inter-brane distance becomes as
short as the local string length, a state of strings connecting the
brane and the antibrane becomes tachyonic and it develops a
non-vanishing expectation value. This ``rolling'' tachyon is similar to
the waterfall direction in a usual field-theoretic hybrid inflation. The
hight of the tachyon potential coincides with the sum of brane tension
and antibrane tension and, according to Sen's
conjecture~\cite{Sen:2002nu}, condensation of the tachyon towards a
potential minimum is an effective description of the brane-antibrane
annihilation.

For a non-vanishing string coupling, the ``rolling'' tachyon couples to 
closed strings. Highly excited closed strings are produced during the
tachyon condensation and the energy stored in the brane-antibrane pair
is rapidly dissipated within the local string time scale. Those excited
closed string modes easily cascade down to massless closed string modes, 
i.e. $10$-dimensional gravitons, again within the local string time
scale. This cascade decay is mainly into Kaluza-Klein (KK) modes rather
than $4$-dimensional gravitons since the $4$-dimensional gravitons are
localized in the bulk Calabi-Yau region and KK modes are localized in
the throat region. Indeed, since the brane-antibrane annihilation takes
place deep inside the warped throat, it is rather natural to expect that
the branching ratio of the decay into the $4$-dimensional gravitons is 
exponentially suppressed compared with that into KK modes. Therefore,
the inflaton throat will be filled with KK gravitons localized in the
throat within the local string time scale after the emergence of the
tachyon.

In the multi-throat scenario, the KK gravitons localized in the inflaton 
throat cannot be the eigen state of the Hamiltonian even in the
quadratic order. This situation is similar to solving Schr\"odinger 
equation with a double well potential. The lowest eigen state in this
quantum mechanical problem is symmetric and the second lowest eigen
state is antisymmetric so that a state localized near one of the two
local minima is a linear combination of the symmetric and antisymmetric
states. Since these two eigen states have different energies, the linear
combination is not an eigen state of the Hamiltonian and, thus, evolves
to different linear combinations of the eigen states as time
proceeds. In other words, the peak of the wavefunction oscillates
between two minima. The time scale of the oscillation is given by the
inverse of the energy difference between the two eigen states. We expect
a similar phenomenon for the KK modes localized in one of throats in 
multi-throat situation: the peak of the KK graviton wave functions should
oscillate among throats~\cite{Dimopoulos:2001ui,Dimopoulos:2001qd}.
When a KK graviton wave function has a peak in the SM throat during the
oscillation, the KK graviton can interact with and decay into the SM fields.

Successful reheating is possible only if the decay into the SM fields
dominates over the decay into the $4$-dimensional gravitons. The former
is suppressed by the relatively long time scale of the oscillation of
the KK graviton wave functions. On the other hand, the latter, which can
happen at any stages of reheating, is suppressed by the warping of the
throats. The result of
refs.~\cite{Barnaby:2004gg,Kofman:2005yz,Chialva:2005zy} indicates that
the decay into the SM field indeed dominates over that into the
$4$-dimensional gravitons for a range of parameters.

In multi-throat scenarios with the inflaton throat and the SM throat, it
is natural to expect that there are more than two throats and that there
are branes in some of them. If there are other branes then light fields
on them are also excited by the above chain of processes. For example,
if there is a throat with a stack of D-branes at the tip then gauge
fields on it will be excited. Moreover, if the initial numbers of branes
and antibranes in the inflaton throat are not the same then there will
be leftover branes or antibranes and gauge fields on them will be
excited by more direct processes. For the success of big-bang nucleo
synthesis, we have to make it sure that more than $90\%$ of energies
distributed over many branes is going into the SM brane.

\subsection{Breakdown of the effective field theory}

A more severe problem is breakdown of the low-energy effective field
theory (EFT) on the SM brane during inflation.

The local string scale in warped compactification depends on the
physical position in extra dimensions. In multi-throat scenarios, since
the warp factor varies from throat to throat, each throat has each local
string scale at the tip. Since one of the main motivations for
considering multi-throat scenarios is to accommodate more than one mass
hierarchies in a natural way, it is plausible to suppose that the scale
of inflation and the electroweak scale are essentially the local string
scales at the tip of the inflaton throat and the SM throat,
respectively. In this case, since the Hubble scale during inflation is
much higher than the electroweak scale, the local string scale at the
tip of the SM throat is expected to be much lower than the Hubble scale
during inflation.

The above argument is based on the picture that the $4$-dimensional
Planck scale is fixed while the local string scale changes as a brane of
interest moves in extra dimensions. In other words, this picture sets
the scale of reference by the universal $4$-dimensional metric $ds_4^2$
in (\ref{eqn:10d-metric}). Alternatively, we could use the metric
$h^2ds_4^2$ to set the scale of reference. In this case the string scale
would be universal everywhere since this metric corresponds to the
induced string metric on the brane. On the other hand, in this picture
the $4$-dimensional Planck scale would change as a brane of interest
moves in extra dimensions. Nonetheless, the ratio of the string scale to
the $4$-dimensional Planck scale at the position of a brane in this
picture should be the same as that in the previous picture. Actually, as
far as dimensionless physical quantities such as ratios among mass
scales are concerned, the two pictures give the same answer as they
should. In this paper we adopt the former picture in which the
$4$-dimensional Planck scale is universal. Note that a mass scale is
dimensionful and, thus, it has physical meaning only when it is compared 
with another mass scale. For example, the ratio of the local string
scale at the position of a brane to the $4$-dimensional Planck scale has
physical meaning irrespective of a scale of reference.

Coming back to the issue of breakdown of the EFT, if the SM brane is at
the tip of the SM throat then stringy effects become dominant on the SM
brane during inflation since the Hubble scale exceeds the local string
scale. This leads to breakdown of the EFT on the SM brane. In this case
we expect violent physical processes such as production of black holes,
strings, topological defects and so on. At the very least, cosmological
density perturbations should significantly deviate from the Gaussian
distribution due to nonlinear natures of the strongly coupled physics
and will most likely contradict with current observation.

\section{A scenario}
\label{sec:scenario}

The aim of this section is to propose a reheating scenario without
breakdown of the EFT on the SM brane.

The position of the SM brane in extra dimensions is a modulus field. At
low energy this modulus field is supposed to be stabilized near the tip
of the SM throat. On the other hand, during inflation, a moduli
potential usually obtain $H$-dependent corrections, where $H$ is the
Hubble scale during inflation~\cite{Dine:1995kz}. For example, a modulus
field can obtain a $H$-dependent mass term and the origin of the
$H$-dependent mass term does not have to be at the tip of a
throat~\footnote{
The derivation of the $H$-dependent mass term via supergravity does not
specify the origin of the $H$-dependent mass term: the dependence of the
K\"ahler potential on the brane position is determined so that it
recovers the correct kinetic term for the brane
position~\cite{DeWolfe:2002nn}, which does not depend on the origin.}. 
Indeed, since a multi-throat configuration is somehow similar to a
double-well potential as discussed in subsection~\ref{subsec:reheating},
it seems natural to expect that the origin of the $H$-dependent mass
term is in or near the bulk region. Thus, it is plausible that the
$H$-dependent corrections to the moduli potential shift the minimum of
the potential for the position of the SM brane. The length of the SM
throat is also a modulus and is expected to be shifted by $H$-dependent
corrections to its potential~\cite{Frey:2005jk}. In the following we
suppose that the modulus corresponding to the brane position is lighter
than the modulus corresponding to the length of the throat. There may be
branes other than the SM brane and the brane-antibrane pair, and their
positions are also moduli. They also start moving when the Hubble scale
becomes comparable with their mass, and oscillate around the potential 
minima. Considering the fact that there is no hierarchal mass scales
between the electroweak scale and today's Hubble scale, we assume that 
there is no moduli with masses in the range between these two scales. In
this case, the SM brane will be the one that started oscillating
last. For this reason, in the following we shall consider the motion of
the SM brane only.

For an EFT to be valid on the SM brane, the local string scale at its
position during inflation must be higher than the Hubble scale and,
thus, much higher than the local string scale at the tip of the SM
throat. This means that the validity of the EFT requires that the
minimum of the potential for the SM brane position during inflation 
should be significantly different from that for $H=0$. This is possible,
for example, if the origin of the $H$-dependent mass term for the brane
position is in or near the bulk Calabi-Yau region. In this case the SM
brane will start moving towards the tip of the SM throat when the Hubble
expansion rate becomes sufficiently low after inflation. Before this
epoch, the energy stored in the brane-antibrane pair is distributed to
many light fields on various branes by the chain of processes explained
in subsection~\ref{subsec:reheating} and is diluted. Eventually, the SM
brane will oscillate around the tip of the
throat~\cite{Mukohyama:2005cv} and fields on the SM brane will be
excited due to interactions with the brane position modulus. This kind
of interaction is in general suppressed only by the local string scale
(but not by the Planck scale) and, thus, the modulus effectively decays
into the SM fields.

Of course, after the SM brane settles down near the tip of the throat,
the modulus field corresponding to the position of the SM brane is
massive and does not affect physics at sufficiently low energies.

\subsection{A toy model: U(1) on wrapped $D5$}

In type IIB superstring theory the SM may be realized on intersecting
$D$-branes. In this subsection, to see that the motion of the SM brane
excites fields on it, we consider a simpler toy model: a wrapped
$D5$-brane and a $U(1)$ field on it. We shall see that the $U(1)$ field
is indeed excited by the homogeneous motion of the $D5$-brane.

The action for a probe $D5$-brane is given as the sum of
Dirac-Born-Infeld and Chern-Simon terms 
%
\begin{equation}
 S_{D5} = -T_5\int d^6\xi e^{-\Phi}
  \sqrt{-\det(G_{AB}-B_{AB}-2\pi\alpha'F_{AB})}
  + T_5\int d^6\xi B\wedge C_4,
\end{equation}
where $\xi^A$ ($A=0,\cdots,5$) are intrinsic coordinates 
on the brane, $T_5$ is the brane tension, $\Phi$ is the dilaton,
$G_{AB}$ is the induced metric on the brane, $B_{AB}$ and $C_4$ are
pullbacks of the NS-NS antisymmetric field and the R-R $4$-form
potential on the brane world-volume, and $F_{AB}$ is the field strength
of the $U(1)$ gauge field on the brane. For simplicity we assume that 
the dilaton is stabilized to zero as in the Klebanov-Strassler
throat~\cite{Klebanov:2000hb} with the GKP construction, and consider
the warped geometry of the form (\ref{eqn:10d-metric}) with 
%
\begin{equation}
 ds_6^2 = dr^2 + ds_5^2, 
\end{equation}
where $ds_5^2$ is an $r$-dependent $5$-dimensional metric.

We consider radial motion of a $D5$-brane wrapped over a $2$-cycle in 
the warped geometry and shall see that the homogeneous motion of the
brane excites the $4$-dimensional part of the $U(1)$ field. We adopt a
gauge in which first four brane coordinates $\xi^{\mu}$
($\mu=0,\cdots,3$) coincide with $x^{\mu}$ and the other brane
coordinates $\xi^p$ ($p=4,5$) coincide with coordinates on the
$2$-cycle, respectively, and suppose that $r$ depends only on
$\xi^{\mu}$. In this case, 
%
\begin{equation}
 G_{AB}d\xi^Ad\xi^B = h^2
  \left(\eta_{\mu\nu} + h^{-4}\partial_{\mu}r\partial_{\nu}r
  \right)d\xi^{\mu}d\xi^{\nu}
  + h^{-2}\sigma_{pq}d\xi^pd\xi^q, 
\end{equation}
where $\sigma_{pq}d\xi^pd\xi^q$ is the pullback of the $5$-dimensional
metric $ds_5^2$ on the $2$-cycle. Thus, we obtain 
%
\begin{equation}
 -\det(G_{AB}-B_{AB}) = 
  h^4(1+h^{-4}\eta^{\mu\nu}\partial_{\mu}r\partial_{\nu}r)(1+B^2)\sigma, 
\end{equation}
where $\sigma\equiv\det\sigma_{pq}$ and $B$ is defined by
%
\begin{equation}
 B_{AB}d\xi^Ad\xi^B = h^{-2}
  B\sqrt{\sigma} \epsilon_{pq}d\xi^pd\xi^q.
\end{equation}
Hereafter, we assume that $h$ and $B$ are functions of $r$ only and
consider a homogeneous brane configuration $r=r(\xi^0)$. Note that
$B_{\mu\nu}$ and $B_{\mu p}=-B_{p \mu}$ have been set to zero in order
to avoid breaking the $4$-dimensional local Lorentz invariance.

As for $F_{AB}$, for simplicity we set $F_{\mu p}=-F_{p\mu}=0$ and 
$F_{pq}=0$. Up to the order $O(F^2)$, this is a consistent truncation
since these components decouple from $F_{\mu\nu}$ in the action up
to this order. We also assume that $F_{\mu\nu}$ depends on the
$4$-dimensional coordinates $x^{\mu}$ only, i.e. we drop massive
Kaluza-Klein modes. This is also a consistent truncation up to the order
$O(F^2)$. Up to the second order in $F_{\mu\nu}$, we obtain 
%
\begin{equation}
 \frac{\sqrt{-\det(G_{AB}-B_{AB}-2\pi\alpha'F_{AB})}}
  {\sqrt{-\det(G_{AB}-B_{AB})}}
  = 1 - \frac{(2\pi\alpha')^2}{2h^4}
  \left( \frac{\delta^{ij}F_{0i}F_{0j}}{1-(\partial_0r)^2/h^4}
   - \frac{1}{2}\delta^{ij}\delta^{kl}F_{ik}F_{jl}\right), 
\end{equation}
where $i,j=1,2,3$. Thus, after integration over the $2$-cycle we obtain
%
\begin{equation}
 S_{D5} = \int d\xi^4
  \left[- \frac{T(\phi)}{\tilde{\gamma}} + T(\phi) - V(\phi)
   + \frac{\delta^{ij}}{2g^2(\phi)}
   \left(\tilde{\gamma}F_{0i}F_{0j}
    - \frac{1}{2\tilde{\gamma}}\delta^{kl}
    F_{ik}F_{jl} \right) \right],
\end{equation}
where
%
\begin{eqnarray}
 T(\phi) & \equiv & T_5v_2h^2\sqrt{1+B^2}, \nonumber\\
 \frac{1}{g^2(\phi)} & \equiv & (2\pi\alpha')^2T/h^4, \nonumber\\
 \tilde{\gamma} & \equiv & \frac{1}{\sqrt{1-(\partial_0\phi)^2/T(\phi)}},
\end{eqnarray}
and we have expressed the contribution from the Chern-Simon term as
$T-V$. Here, $\phi$ is a field defined by
%
\begin{equation}
 d\phi = \left(T_5v_2h^{-2}\sqrt{1+B^2}\right)^{1/2}dr,
\end{equation}
and $v_2$ is the (unwarped) volume of the $2$-cycle
%
\begin{equation}
 v_2 \equiv \int d\xi^4d\xi^5\sqrt{\sigma},
\end{equation}
which depends on $\phi$.

Now, assuming that the physical energy scale associated with the
$4$-dimensional universe is much lower than the energy scale of moduli
stabilization, we promote the $4$-dimensional effective action to the
FRW background. By replacing $\eta_{\mu\nu}d\xi^{\mu}d\xi^{\nu}$ with 
$a^2(\xi^0)\eta_{\mu\nu}d\xi^{\mu}d\xi^{\nu}$ and repeating the 
above procedure, we obtain $S_{D5}=S_{\phi}+S_A$, where
%
\begin{eqnarray}
 S_{\phi} & = & \int d\xi^4\ a^4
  \left[- \frac{T(\phi)}{\gamma} + T(\phi) - V(\phi)\right],
  \nonumber\\
 S_A & = & \int d\xi^4
  \frac{\delta^{ij}}{2g^2(\phi)}
   \left(\gamma F_{0i}F_{0j}
    - \frac{1}{2\gamma}\delta^{kl}F_{ik}F_{jl} \right),
\end{eqnarray}
where
%
\begin{equation}
 \gamma \equiv \frac{1}{\sqrt{1-a^{-2}(\partial_0\phi)^2/T(\phi)}}
\end{equation}

Note that both $S_{\phi}$ and $S_A$ should in general receive
corrections due to moduli stabilization, radiative corrections,
curvature corrections and so on. However, the $U(1)$ gauge symmetry and
the spatial rotational invariance protect the form of $S_A$. (The form
of $\gamma$ and $g^2(\phi)$ may change.) What is important in the
following analysis is the form of $S_A$ only.

Let us now see that the $U(1)$ field should be excited by the homogenous
motion of the brane. This is similar to the decay of a moduli field into
gauge field via the dependence of the gauge coupling on the moduli
field. By setting $A_0=0$ and $\delta^{ij}\partial_iA_j=0$, we obtain
%
\begin{equation}
 S_A = \int d\xi^4
  \frac{\delta^{ij}}{2g^2(\phi)}
   \left(\gamma\partial_0A_i\partial_0A_j
    - \frac{\delta^{kl}}{\gamma}\partial_kA_i\partial_lA_j\right),
\end{equation}
and the corresponding equation of motion is
%
\begin{equation}
 \gamma g^2\frac{d}{d\eta}
  \left(\frac{\gamma}{g^2}\frac{dA_i}{d\eta}\right)
  - \delta^{kl}\partial_k\partial_lA_i = 0,
\end{equation}
where $\eta=\xi^0$. During inflation, $\phi$ stays almost constant and,
thus, $\gamma\simeq 1$ and $g^2$ is almost constant. Thus, $A_i$ is
expanded as 
%
\begin{equation}
 A_i = \int \frac{d^3\vec{k}}{\sqrt{(2\pi)^3}}
  \sum_{I=1,2}e^I_i(\vec{k})
  \left[ a_I(\vec{k})\varphi(\vec{k},\eta)e^{i\vec{k}\cdot\vec{\xi}}
   +a^{\dagger}_I(\vec{k})\varphi^*(\vec{k},\eta)e^{-i\vec{k}\cdot\vec{\xi}}\right],
\end{equation}
where $e^I_i$ ($I=1,2$) is the polarization vector satisfying 
%
\begin{equation}
 e^I\cdot e^J = \delta^{IJ}, \quad e^I\cdot\vec{k} = 0, 
\end{equation}
and $\varphi$ satisfies 
%
\begin{equation}
 \varphi \simeq \frac{g_{inf}}{\sqrt{2|\vec{k}|}}e^{-i|\vec{k}|\eta}
\end{equation}
during inflation, and 
%
\begin{equation}
 \gamma g^2\frac{d}{d\eta}
  \left(\frac{\gamma}{g^2}\frac{d\varphi}{d\eta}\right)
  + |\vec{k}|^2\varphi = 0
\end{equation}
for all the time since then. Here, $g_{inf}$ is the (almost constant)
value of $g$ during inflation. Long after inflation, the modulus $\phi$
settles down to the value at the minimum of its potential and, thus,
$\gamma\simeq 1$ and $g^2$ approaches a constant. Thus, $A_i$ is
expanded as 
%
\begin{equation}
 A_i = \int \frac{d^3\vec{k}}{\sqrt{(2\pi)^3}}
  \sum_{I=1,2}e^I_i(\vec{k})
  \left[ \tilde{a}_I(\vec{k})\tilde{\varphi}(\vec{k},\eta)
   e^{i\vec{k}\cdot\vec{\xi}}
   +\tilde{a}^{\dagger}_I(\vec{k})\tilde{\varphi}^*(\vec{k},\eta)
   e^{-i\vec{k}\cdot\vec{\xi}}\right],
\end{equation}
where $\tilde{\varphi}$ satisfies 
%
\begin{equation}
 \tilde{\varphi} \simeq \frac{g_0}{\sqrt{2|\vec{k}|}}e^{-i|\vec{k}|\eta}
\end{equation}
long after inflation, and
%
\begin{equation}
 \gamma g^2\frac{d}{d\eta}
  \left(\frac{\gamma}{g^2}\frac{d\tilde{\varphi}}{d\eta}\right)
  + |\vec{k}|^2\tilde{\varphi} = 0
\end{equation}
until then. Here, $g_0$ is the asymptotic value of $g$ at late
time. Since $g^2$ is time-dependent between the two asymptotic regimes,
the two sets of mode functions $\varphi$ and $\tilde{\varphi}$ are not
identical. (The time dependence of $\gamma$ can be absorbed by
redefinition of the time variable.) Instead, they are related as 
%
\begin{equation}
 \tilde{\varphi}(\vec{k},\eta) = \alpha(\vec{k})\varphi(\vec{k},\eta)
  + \beta(\vec{k})\varphi^*(\vec{k},\eta). 
\end{equation}
Hence, if we start with the initial vacuum defined by
%
\begin{equation}
 a_I(\vec{k})|in\rangle = 0 \quad 
  \mbox{ for }{}^{\forall}I\mbox{ and }{}^{\forall}\vec{k},
\end{equation}
then the number of created particles at late time is
%
\begin{equation}
 \langle in|\tilde{a}_I^{\dagger}(\vec{k})\tilde{a}_I(\vec{k})|in\rangle
  = |\beta(\vec{k})|^2 \ne 0.
\end{equation}
This shows that the motion of the wrapped $D5$-brane excites the $U(1)$
field on it.

The above analysis explicitly shows that the neutral modulus field
$\phi$ can directly decay into the $U(1)$ gauge field. Up to the order
$O(F_{\mu\nu}^2)$, the decay is due to the coupling of the modulus field
to the gauge field via the dependence of the gauge coupling on the
modulus field. On the other hand, in higher order, there are many other
decay channels although we probably have less theoretical controls. In
this sense, what we have seen above is just one of many possible
processes to excite the $U(1)$ field by the brane motion.

After all, oscillation of the SM brane is rather violent phenomenon: the 
brane loading all components of the SM, including ourselves, is moving!
It is not at all unreasonable to expect that excitation of the SM fields
due to the brane motion is efficient.

\subsection{Reheating temperature}

As we have seen in the previous subsection, the SM fields can be excited
by motion of the SM brane. Processes include
non-perturbative~\cite{Kofman:1997yn} as well as perturbative decays of
the modulus field. In this subsection, we shall estimate the reheating
temperature based on the decay rate in the perturbative regime (with
small amplitude of the oscillation of $\phi$). Processes in the
non-perturbative regime (with large amplitude of the oscillation of
$\phi$) are certainly worthwhile investigating as a future work.

In scenarios with warped extra dimensions, it is expected that
interaction between the modulus corresponding to the brane position and
fields on the brane is suppressed not by the Planck scale but by the
local string scale. To see this let us suppose that at low energy, a
$D5$-brane wrapped over a $2$-cycle is stabilized somewhere, say at 
$r=r_0$, in a warped throat. We assume that the warp factor at $r=r_0$
is small enough to address the hierarchy problem. Because of the warped
geometry, dimension-full quantities on the brane should scale as some
powers of the warp factor. In particular, unless fine-tuned, we expect
that 
%
\begin{equation}
 \frac{d\ln g}{d\phi} \sim \frac{C(g_s)}{M_{local}},
\end{equation}
where $M_{local}$ is the local string scale defined by  
%
\begin{equation}
 M_{local} \equiv \frac{h_{tip}}{\sqrt{\alpha'}}.
  \label{eqn:Mlocal}
\end{equation}
(Note that $\phi$ has been normalized so that its leading kinetic term
is canonical.)

Thus, Lagrangian is expanded near the present value $\phi=\phi_0$ as 
%
\begin{equation}
 -\frac{1}{4g_0^2}
  \left[F^{\mu\nu}F_{\mu\nu}
   + \frac{O(1)}{M_{local}}(\phi-\phi_0)F^{\mu\nu}F_{\mu\nu}
   + \cdots \right].
\end{equation}
Therefore, in this toy model the interaction between the modulus $\phi$
and the $U(1)$ gauge field is suppressed by the local string scale. Note
that the warp factor $h$ at $r=r_0$ is assumed to be exponentially small
(see (\ref{eqn:warp-factor}))  so that we have the hierarchy 
%
\begin{equation}
 M_{local} \ll M_{Pl}, 
\end{equation}
where $M_{Pl}$ is the $4$-dimensional Planck scale given by 
%
\begin{equation}
 M_{Pl}^2\simeq \frac{2V_6}{(2\pi)^7{\alpha'}^4g_s^2}.
\end{equation}
Here, $V_6$ is the volume of the bulk Calabi-Yau region.

Since the suppression of the moduli interaction is just by the local
string scale,  it is expected that the decay rate
of $\phi$ into the SM fields in the perturbative regime is 
%
\begin{equation}
 \Gamma \sim \frac{m_{\phi}^3}{M_{local}^2},
\end{equation}
where the local string scale $M_{local}$ is given by
(\ref{eqn:Mlocal}). Correspondingly, the reheating temperature is
estimated as 
%
\begin{equation}
 T_{reh} \sim \frac{\sqrt{M_{Pl}m_{\phi}^3}}{M_{local}}. 
\end{equation}
Since $M_{Pl}$ is not in the denominator, it is easy to make the
reheating temperature sufficiently high.

\section{Summary}
\label{sec:summary}

More than one mass hierarchies can coexist in multi-throat scenarios of
warped extra dimensions. A long throat can accommodate a brane confining
the standard model (SM) of particle physics while another throat with an 
intermediate length can accommodate a brane-antibrane pair to drive
inflation. In this paper, after pointing out a couple of problems with
the multi-throat inflation scenario, we have proposed a new reheating
mechanism in which motion of the SM brane excites fields on
it. Since interaction is suppressed only by the local string scale, the
modulus corresponding to the brane position effectively decays into the
SM fields. We have considered a wrapped $D5$-brane as a simple toy model
of the SM brane and show that the brane motion indeed excites the $U(1)$
gauge field on it. The decay process that we have explicitly
investigated is similar to the decay of a moduli field into a gauge
field via moduli dependence of the gauge coupling, but in general there
are many other decay channels.  

In some sense, this is an illustration how the moduli problem reappears
and is solved in brane world scenarios with warped extra dimensions. In 
the present case, interaction between the moduli field and the SM fields
is suppressed not by the Planck scale but by the local string scale at
the position of the SM brane. Since the local string scale is much lower
than the Planck scale, the moduli effectively decays and reheats the
universe. Positions of other branes are also moduli but they decay
before the SM brane, provided that they are heavier than the electroweak
scale as discussed in the beginning of Sec.~\ref{sec:scenario}.

In order to estimate reheating temperature, it is important to consider
more realistic models of the SM brane such as intersecting $D$-branes
and investigate how the brane motion and fields on it interact with each
other. It is also worthwhile understanding how this mechanism of
reheating can be understood in terms of scalar-tensor theory.

\section*{Acknowledgements}

The author would like to thank J.~Yokoyama for helpful discussions and
L.~Kofman for useful comments. This work was in part supported by MEXT
through a Grant-in-Aid for Young Scientists (B) No.~17740134 and by JSPS
through a Grant-in-Aid for Creative Scientific Research No.~19GS0219 and
through a Grant-in-Aid for Scientific Research (B) No.~19340054.  


\end{document}